\RequirePackage{fix-cm}
\documentclass[10pt,journal,compsoc]{IEEEtran}
\usepackage{amsmath}
\usepackage{graphicx}
\usepackage{accents}
\usepackage{fixltx2e}
\usepackage{tabularx}
\usepackage[boxed,linesnumbered]{algorithm2e}
\usepackage{xcolor}
\usepackage{url}
\usepackage{color}

\SetAlFnt{\small}
\SetAlCapFnt{\small}
\SetAlCapNameFnt{\small}
\SetAlCapHSkip{0pt}
\IncMargin{-\parindent}

\newcommand\crule[3][black]{\textcolor{#1}{\rule{#2}{#3}}}
\definecolor{Green}{rgb}{0,0.5,0}
\definecolor{Magenta}{rgb}{1,0,1}
\definecolor{cherryblossompink}{rgb}{1.0, 0.72, 0.77}
\definecolor{darkseagreen}{rgb}{0.56, 0.74, 0.56}
\definecolor{etonblue}{rgb}{0.59, 0.78, 0.64}
\definecolor{flamingopink}{rgb}{0.99, 0.56, 0.67}
\definecolor{myblue}{RGB}{175,175,200}
\definecolor{mywhite}{RGB}{215,215,215}

\def\scm {{\sf $M_{sc}$}}
\def\sm {{\sf $M_m$}}

\newcommand{\ie}{\unskip, \textit{i.\,e.},\ }

\ifCLASSOPTIONcompsoc
  \usepackage[nocompress]{cite}
\else
  \usepackage{cite}
\fi
\ifCLASSINFOpdf
\else
\fi
\begin{document}


\title{Use of Source Code Similarity Metrics in Software Defect Prediction}

\author{Ahmet Okutan        
\IEEEcompsocitemizethanks{\IEEEcompsocthanksitem Ahmet Okutan is with the Department of Computer Engineering, Rochester Institute of Technology, Rochester, NY, USA.\protect}
\thanks{}}

\IEEEtitleabstractindextext{%
\begin{abstract}
In recent years, defect prediction has received a great deal of attention in the empirical software engineering world. Predicting software defects before the maintenance phase is very important not only to decrease the maintenance costs but also increase the overall quality of a software product. There are different types of product, process, and developer based software metrics proposed so far to measure the defectiveness of a software system. This paper suggests to use a novel set of software metrics which are based on the similarities detected among the source code files in a software project. To find source code similarities among different files of a software system, plagiarism and clone detection techniques are used. Two simple similarity metrics are calculated for each file, considering its overall similarity to the defective and non defective files in the project. Using these similarity metrics, we predict whether a specific file is defective or not. Our experiments on 10 open source data sets show that depending on the amount of detected similarity, proposed metrics could achieve significantly better performance compared to the existing static code metrics in terms of the area under the curve (AUC).
\end{abstract}

}

\maketitle

\section{Introduction}
\label{introd}
Testing is regarded as one of the most time and resource consuming phases in the software engineering life cycle, because underestimating tests can lead to serious problems for customers during the maintenance. According to Brooks, considering component and system tests together, testing and debugging activities takes about 1/2 of the project time, while coding takes only about 1/6 of it \cite{Brooks:1995}. On the other hand, predicting defects in early development stages or at least before the maintenance is very crucial for project schedule, budget, and quality. The later a defect is observed and fixed means more cost and less customer satisfaction in a software project. Fagan states that the cost of reworking software bugs in programs becomes higher the later they are reworked in the software development process, so every attempt should be made to find and fix errors as early as possible in the development process \cite{Fagan1976}. 

Software defect prediction helps software engineers to improve the efficiency of the testing phase and gives the opportunity to increase the overall quality of a software product. According to Boehm and Papaccio, software rework tends to follow the Pareto-Zipf-type law, where 80 percent of the software rework costs result from 20 percent of problems \cite{Boehm:1988}. Therefore, while allocating resources during the testing period, project managers should make use of the defect prediction techniques to make more optimal decisions and focus on the more defect prone modules rather than the less or non defective ones.

Fenton and Neil state that one of the key issues affecting the historical research direction of the software engineering community is the problem of using size and complexity metrics as sole predictors of defects \cite{Norman03}. Different and novel software metrics are always welcome, especially if their defect prediction performance is comparable or better on certain experiment contexts when compared to the existing metrics. As the software projects get larger and the amount of the shared code pieces increases within a project, the importance of measuring inter file attributes increases as well. This paper proposes a novel set of software metrics that are based on source code similarities detected among the source code files of a project and shows that proposed metrics are better defect predictors compared to the traditional source code metrics. 

Software metrics play a key role in measuring the defect proneness of a software product since the usefulness of a defect prediction model is highly dependent on the the quality of the underlying software metrics. Different types of software metrics have been proposed so far to measure software defect proneness. Software metrics are divided into two main categories as product and process based metrics. According to Henderson-Sellers \cite{Henderson01} product metrics take the snapshot of a software at a certain time, whereas process metrics measure the software changes over time. Some of the product based metrics that measure the source code related attributes of a software product are Object Oriented metrics (CK suite) \cite{Chidamber01}, Halstead metrics \cite{Halstead01}, McCabe metrics \cite{McCabe02}, and line of code metrics. On the other hand, process based metrics that measure the changing nature of the software are collected over a certain period of time and might be related to developers, revisions and source code changes. For example number of developers  \cite{Schroter06} \cite{Ostrand10} \cite{Weyuker01} \cite{Graves01} \cite{okutan2} \cite{Krishnan2011} \cite{Moser:2008}, number of revisions \cite{PurushothamanP05} \cite{Krishnan2011} \cite{Moser:2008}, and source code change related metrics \cite{NagappanB07} \cite{Graves01} \cite{Krishnan2011} \cite{Krishnan2011} \cite{Shin11} \cite{Moser:2008} are some of the process metrics that are commonly used in the defect prediction literature.

Static code metrics have a very long history in the defect prediction literature considering the fact that the Halstead \cite{Halstead01} and McCabe  \cite{McCabe02} metrics were defined in 1970s. Furthermore, the extraction of static code metrics is relatively easy, because there are automated tools that mine the source code repositories of projects and calculate static code measures in a feasible amount of time. However, most of the static code metrics proposed so far show intra module characteristics \ie measure inner module attributes. Although some of them take into account inter module relationships \cite{Chidamber01}, they can be used only if the object oriented programming paradigm is used. The metrics proposed in this paper are based on the source code similarities among the files of a software project, and take into account the inter module (file) relationships independent from the underlying software development methodology. This is a key novelty of the metric set defined in this paper, when compared to the existing source code metrics in the software defect prediction literature. 

In this paper, the relationship between the syntactic or semantic similarities of source files and their defect proneness is investigated. Based on the hypothesis that a defective file could be more similar to a defective file, when two source code files are similar, we wonder if there is a relationship between their similarity and defectiveness? To measure the similarity among different source code files, the state of the art plagiarism and clone detection techniques are used. After finding the similarities among the source code files of a software system, two source code metrics are calculated for each file. One of these metrics is named Similarity to Defectives (STD) and used to measure the overall similarity of a file to all defective files in the software system. The second metric is named Similarity to Non-defectives (STN) and used to calculate the overall similarity of a file to all non defective files in the system. We use these novel source code similarity metrics that are summarized in Section \ref{metric_calculation_details} in order to predict software defectiveness. We test the performance of the proposed similarity metrics on 10 open source data sets listed in Table \ref{data_sets}. The area under the ROC curve (AUC) is used as performance indicator, to show that the performance of the proposed metrics is significantly better when compared to the existing static code metrics listed in Table \ref{static_code_metrics}.

This paper is organized as follows: In Section \ref{previous_work}, we present a brief review of the related work about software metrics and source code similarity detection in the literature. We explain the proposed source code similarity metrics in Section \ref{Proposed_Approach}. We describe the experiment methodology and discuss our results in Section \ref{Experiments_and_Results} before we conclude in Section \ref{Conclusion}.

\section{Previous Work}
\label{previous_work}
While developing complex software applications composed of thousands of lines of source code, the probability of committing errors is quite high. Therefore, no software system of any realistic size is ever completely debugged \cite{Yourdon}. This research proposes to use a novel metric set that is based on the similarity of the source files in a software system to predict software defectiveness. Before describing the details of the proposed methodology, we provide a summary of the previous studies in defect prediction, software metrics, and different source code similarity detection approaches.

Hall et al. \cite{Hall2012} carry out a systematic literature review using 208 fault prediction studies published between January 2000 and December 2010, and find out that simple modelling techniques like naive Bayes or logistic regression perform better than more complex modelling techniques like SVM. They conclude that the methodology used to build a prediction model seems to be significant for predictive performance. Lessmann et al. \cite{Lessmann2008} use 10 public domain data sets from the NASA Metrics Data repository to compare the defect prediction performance of 22 classifiers. They conclude that the importance of a specific classification technique is less important than it was assumed to be before, because the differences in the prediction performance of the top 17 classifiers are not statistically significant. On the other hand, Song et al. \cite{Song2011} propose a general framework for software defect prediction to analyze the prediction performance of competing learning schemes and conclude that it is better to choose different learning schemes for different data sets since small details in the evaluation of experiments can completely change findings. Shepperd and Kadoda \cite{Shepperd01} state that there is a strong relationship between the success of a particular defect prediction technique and characteristics of the underlying data set, so the defect prediction methods must be evaluated in their experiment contexts before saying a technique is the best in general. 

On the other hand, D'Ambros et al. \cite{DAmbros2012} state that external validity is a challenging problem in the defect prediction literature, because although some defect prediction techniques perform better than others in specific contexts, no baseline exists to compare existing prediction approaches. Menzies et al. \cite{Menzies01} state that debates for the merits of different metric sets are irrelevant, because how the metrics are used to build the prediction model is much more important than which particular metric set is used. Therefore, conclusions regarding the best metric(s) may not apply when different data sets are tried. They also conclude that data mining static code metrics to learn defect predictors is still useful. The novel metric set proposed in this research helps to build better prediction models, when the amount of similarity among the files of a software system is high.

\subsection{Software Metrics}
\label{studies_sw_metrics}
Different types of defect prediction metrics exist to provide objective, reproducible and quantifiable measurements about a software product. Daskalantonakis categorize software metrics based on their intended use as product, process and project metrics. He states that although a metric can belong to more than one category, product metrics are used to measure a software product itself, process metrics are used for improving the software development and maintenance process and the project metrics are utilized for tracking and improving a project \cite {Daskalantonakis}. 

McCabe shows that a software module can be modelled as a directed flow graph where each node is for a program statement and each arc is for the flow of control from one statement to another. McCabe introduces the cyclomatic, design and essential complexity terms and suggests to calculate static code metrics considering the directed flow graph generated from a software module \cite{McCabe02}.

Halstead states that software modules that are hard to read tend to be more defect prone. He calculates software complexity of a module by defining four key attributes as the number of operators, the number of operands, the number of unique operators, and the number of unique operands, then derives his metrics that provide insight about the source code complexity \cite{Halstead01}. Although the theory behind Halstead metrics has been questioned repeatedly \cite{Fenton:1998}, together with McCabe metrics, these metrics are one of the most widely known source code complexity metrics in the defect prediction literature.  

Chidamber and Kemerer (CK) propose a metric set specifically for software systems that are developed using the the object-oriented programming paradigm \cite{Chidamber01}. These metrics measure the degree of coupling and cohesion among the classes of a software system, as well as the depth of inheritance relationships. Henderson-Sellers proposes to use LCOM3 metric as an alternative to measure the class cohesion by taking into account the attribute method correlation \cite{Henderson01}. Basili et al. use the object oriented metrics proposed by CK and mark some of them more useful compared to the remaining ones \cite{Basili:1996}. Furthermore Gyimothy et al. compare the accuracy of CK metrics together with some other metrics, to predict defective classes in Mozilla and conclude that CBO has the best predictive power \cite{Gyimothy01}. Okutan and Yildiz use CK object oriented metric set together with line of code metric and show that response for class (RFC) is one of the most effective metrics on defectiveness, whereas the number of children (NOC) and the depth of inheritance tree (DIT) have very limited effect and are untrustworthy \cite{okutan2}.  

Bansiya and Davis state that there are metrics to model design properties like abstraction, messaging and inheritance, but no object oriented design metrics exists for several other design properties like encapsulation and composition. Furthermore, they state that some existing metrics to measure complexity, coupling and cohesion require almost a full implementation and that is why are difficult to generate during design. So, they define five new metrics (known as QMOOD metrics suite) which can be calculated from the design information only \cite{Bansiya01}.

Line of code metrics are one of the most widely used static code metrics to measure the size of a software system. The importance of each single line could be different especially when considering whether it is an empty, comment or a single bracket (or begin) line. Therefore, different types of line of code counts proposed to measure each of these different types of lines like loc\_total, loc\_blank, loc\_code\_and\_comment, loc\_comments, and loc\_executable. 

In addition to the static code, process, and line of code metrics, there are some other types of measures that are used to predict software defects. Zimmermann et al. propose a prediction model that is based on Windows Server 2003 to generate a module dependency graph and conduct network analysis on that graph. They define network metrics like degree of centrality, closeness, and betweenness centrality and show that network metrics on dependency graphs are useful to predict the number of defects. Furthermore, they also show that network metrics are better in predicting critical binaries and defectiveness when compared to the complexity metrics \cite{Zimmermann:2008}. Meneely et al. use a new set of developer metrics that are based on a developer social network where two developers are connected if they work on the same files during the same period of time. They observe a significant correlation between the file based developer metrics and software failures \cite{Meneely:2008}.

\subsection{Source Code Similarity}
There are previous research studies about source code clones and plagiarism, but there is no prior study where source code similarity based metrics are used to predict software defectiveness. Below we briefly summarize different techniques that are used in the previous studies to find source code similarity or plagiarism:

\begin{enumerate}

\item \textit{String based:} String based techniques use a simple and straight forward approach where each statement in the program is represented as a string and then the whole program is mapped to a sequence of strings. In order to compare two code fragments, their corresponding sequence of strings are compared. 

\item \textit{Metric based:} Metric based techniques collect certain attributes of the source code and use these attributes to compare different source code files. Kontogiannis et al. use two pattern matching techniques based on source code metrics that characterize source code fragments and a dynamic programming algorithm that allows statement level comparison of feature vectors that represent program statements. They show that proposed pattern matching techniques are useful in detecting similarities between code fragments \cite{Kontogiannis:1996}.  

\item \textit{Tree based:} In this technique, parse trees are generated from the source code and they are compared to find similarities among different code fragments. Baxter et al. propose to use abstract syntax trees to find source code clones. They try their technique on a production software system around 400K lines of code with 19 subsystems and show that the proposed technique is able to find code clones in each sub system \cite{Baxter:1998}. Wahler et al. propose a new approach to detect type 1 and type 2 clones using an an abstract syntax tree representation of the source code in XML. They use the the Java Development Kit (JDK) source repository and a couple of private projects to show the feasibility of their approach \cite{Wahler:2004}.

\item \textit{Token based:} In Token based approaches, the statements in the source code are tokenized and the source code is represented as a sequence of tokens. Then, sub sequences of token clones are searched to find source code similarities. Token based approaches have some advantages over both string and tree based approaches. First of all string based techniques can not tolerate identifier renaming. Second, with tree based approaches having identical abstract syntax trees does not mean that the corresponding code fragments are clones since the variables are disregarded during tree generation. Kamiya et al. propose a tool named CCFinder that uses a token based technique to detect code clones in the C, C++, Java, COBOL source files. They use JDK, FreeBSD, Linux, NetBSD, and several other projects from the industry to show that the their optimized token based comparison technique is successful in detecting code clones \cite{Kamiya1019480}. Li et al. propose a tool named CP-Miner that is based on the token based approach to find source code clones. They show that compared to the CCFinder it finds between 17\% and 52\% more copy pasted source code, since it tolerates code modifications better \cite{Li1610609}.

\item \textit{Semantic:} Semantic techniques use the program dependency graphs (PDG) to detect similarities between code fragments \cite{Liu:2006}. They tolerate statement insertion or reordering and take into account semantic similarities (and data flow) between code fragments. They are less fragile to statement reordering and code insertion compared to other techniques, especially if experienced plagiarists try to disguise. In our research context, the focus is on detecting source code similarities rather than catching any disguises or plagiarised code fragments. Therefore, we use token based similarity detection tools during our experiments.

\end{enumerate}

\section{Proposed Approach}
\label{Proposed_Approach}
We believe that source code similarity based software metrics could be used to predict software defectiveness. Plagiarism detection tools use fingerprint based or content comparison techniques to find structural similarity among source code pieces. We use plagiarism detection tools or clone detection techniques to extract source code similarities and generate a similarity matrix that shows the extent of similarity among the files in a software system. Then using this similarity matrix, we calculate the proposed source code similarity based software metrics for each file in the system.

\subsection{Source Code Similarity Detection}
\label{Finding Source Code Similarities}
Token based similarity detection techniques are more successful compared to the string and tree based approaches. Because, they tolerate both variable renaming and even statement reordering by using fingerprinting \cite{Moss01}. JPlag \cite{Prechelt01} and MOSS \cite{Moss01} are two famous token based tools that are used worldwide to measure source code similarities. JPlag uses a token based approach to extract structural source code similarities and the similarity is calculated based on the percentage of token strings covered. JPlag is not so successful in finding all similarities in a software project due to some parse deficiencies \cite{Cosma01}. MOSS (Measure of Software Similarity) is a token based plagiarism detection tool that tolerates statement insertion or reordering fragility by using fingerprinting. It uses an algorithm that divides programs into $k$-grams, where a $k$-gram is a contiguous substring of length $k$. Each $k$-gram is hashed and a subset of these hashes are selected as the fingerprint of a file. Similarity between any two files is determined by the number of fingerprints shared between these files. MOSS uses an efficient and scalable winnowing algorithm that includes positional information of the fingerprints and tends to use fewer fingerprints compared to previous techniques. Schleimer at al. report that after years of service, there was no false positives and the rare reports of false negatives (instances of common sub strings that could not be discovered) have found to be either an implementation bug or user error \cite{Schleimer2003}. MOSS is a web based tool developed by Alex Aiken and hosted by Stanford University. One can submit the source code of a software project to MOSS (or a collection of projects together) to check plagiarism online \cite{Moss01}. Due to its ease of use and proven successful track record, we use MOSS to extract source code similarities among the files of the software projects used in the experiments. 

Besides the plagiarism detection tools, clone detection techniques are used as an alternative way of finding source code similarities. Karp-Rabin string matching algorithm compares the hash values of string tokens rather than the strings themselves. The copy paste detector (CPD) of the open source PMD source code analyzer uses Karp-Rabin string matching algorithm to find code clones among source code files in a software project \cite{karp87}. Furthermore, CCFinder \cite{Kamiya1019480} uses a token based approach to detect code clones among the source code files. Ragkhitwetsagul et al. state that highly specialised source code similarity detection techniques and tools are more successful compared to the textual similarity measurement approaches and they show that CCFinder is one of the best performing clone detection techniques \cite{Ragkhitwetsagul}. In addition to MOSS, CCFinder and CPD are used as clone detection techniques to find the clone based source code similarities among the source code files of each project.

As a summary, to find similarities among the source code files of each project in Table \ref{data_sets}, MOSS plagiarism detection tool, and CPD and CCFinder clone detection tools are used. The parameters used with each each of these similarity detection tools are described below.
\begin{enumerate}
\item \textbf{MOSS: } MOSS uses two parameters \ie $m$ that represents the maximum number of times a shared code may appear across files and $n$ that determines the maximum number of matching files to include in the results. Assuming there are $N$ files in a project, we use $m = N$ and $n = N^2$ to detect and include all possible similarities. 
\item \textbf{CPD: } The open source PMD source code analyzer scans source code written in Java and other languages and looks for potential problems like bugs, dead or sub-optimal codes, over-complicated code pieces and duplicated code. The Copy/Paste Detector (CPD) of PMD was used to find the number of tokens shared among different files in each project. To detect more similarities, the minimum number of tokens to report was set to 50. The total number of shared tokens between two files is used as a measure of similarity between them. 
\item \textbf{CCFinder: } The ccfx command line utility of the CCFinder clone detection tool is used to extract shared clones among source code files in each project. Ccfx is run in the detection mode (with parameter $-d$) for each project directory to find clones, and then used in the print mode (with parameter $-p$) to print the contents of the the binary clone file generated in the detection mode. 

\end{enumerate}

\subsection{Proposed Similarity Metrics}
\label{metric_calculation_details}
Assuming there are $N$ files in a software system where each file is represented with $f_i$, we generate a similarity matrix $S$ of size $N$ by $N$ that shows the source code similarities among the files. Each $S(f_i,f_j)$ term in the matrix gives the amount of similarity between files $f_i$ and $f_j$.

\begin{equation}
S=
\label{Eq:smatrix}
\begin{bmatrix}
 &S(f_1,f_1)  &S(f_1,f_2)  &...  &S(f_1,f_N) \\ 
 &S(f_2,f_1)  &S(f_2,f_2)  &...  &S(f_2,f_N) \\ 
 &...  &...  &...  &... \\ 
 &S(f_N,f_1)  &S(f_N,f_2)  &...  &S(f_N,f_N) 
\end{bmatrix}
\end{equation}
Then for each file $f_i$ we calculate $STD_i$ (similarity to defective files) and $STN_i$ (similarity to the non defective files) metrics. We find 
\begin{equation}
STD_i = \sum_{j=1}^{j=N} {S(f_i,f_j)}
\end{equation}   
where we take into account $S(f_i,f_j)$ if $f_j$ is defective. Similarly, we calculate 
\begin{equation}
STN_i = \sum_{j=1}^{j=N} {S(f_i,f_j)}
\end{equation}
where we include $S(f_i,f_j)$ in the sum only if $f_j$ is non defective.

On the other hand, we wonder how defectiveness is affected when the file size is considered together with the similarity. For instance, we want to see whether being 20\% similar to a file of size 100 kilobytes is more meaningful when compared to being 20\% similar to a file of size one kilobytes. That is why we introduce two more metrics and compare if they are better defect predictors compared to STD and STN. Assuming the size of a file $f_i$ is $s_i$ in terms of kilobytes, we calculate $STDS_i$ (similarity to defective files with size) and $STNS_i$ (similarity to the non defective files with size) metrics for each file $f_i$ as 
\begin{equation}
STDS_i = \sum_{j=1}^{j=N} {S(f_i,f_j) * s_j}
\end{equation}   
where we take into account $S(f_i,f_j)$ if $f_j$ is defective. Similarly, we calculate 
\begin{equation}
STNS_i = \sum_{j=1}^{j=N} {S(f_i,f_j) * s_j}
\end{equation}
where we include $S(f_i,f_j)$ in the sum only if $f_j$ is non defective. As a summary, the list of the studied source code similarity based software metrics is shown in Table \ref{similarity_metrics_table}.

\begin{table}[h!]
\centering
\caption{List of the Proposed Similarity Metrics.\label{similarity_metrics_table}}{
\begin{tabular}{ll}
\hline
\multicolumn{ 2}{c}{\textbf{List of Similarity Metrics Used}} \\ \hline \hline
\textbf{STD} & Similarity to Defective Files \\ 
\textbf{STN} & Similarity to Non Defective Files \\ 
\textbf{STDS} & Similarity to Defective Files with Size \\ 
\textbf{STNS} & Similarity to Non Defective Files with Size \\ \hline
\end{tabular}}
\end{table}

The density of the similarity matrix ($S$) depends on the extent of the similarity of files in each project (data set). Therefore, the amount of detected similarity among the files of a project is very important while calculating the proposed similarity metrics. For each project, we first use MOSS plagiarism tool to detect the similarities of files, generate a similarity matrix, and calculate the proposed similarity metrics. To cross check our findings, we also use CPD clone detection technique and CCFinder to find the similarities of files, generate alternative similarity matrices, and calculate the same metrics for each project. As a summary, we calculate the STD and STN similarity metrics for MOSS, CPD, and CCFinder in each project.

\section{Experiments and Results}
\label{Experiments_and_Results}

\subsection{Metric Generation Process}
\label{Metric Extraction Process}
To extract similarity metrics listed in Table \ref{similarity_metrics_table} based on the similarities detected by MOSS (with $m = N$, $n = N^2$), CPD (with minimum number of tokens set to 50), and CCFinder (with $-d$ parameter), a special software was developed in Java. The main steps of the metric extraction process are:
\begin{enumerate}
\item First we find the syntactic or semantic similarities among the source code files of each data set using MOSS. The similarity output for each project is saved as a comma-separated file with three columns \ie the name of the first file, the name of the second file and the amount of similarity detected between them. The output file contains all similarities detected between any pairs of source code files in the project. 
\item Using the output in the previous step, the similarity matrix $S$ defined in Equation \ref{Eq:smatrix} is generated for each data set.
\item Based on the similarity matrix $S$, the proposed similarity metrics STD and STN are calculated for each file in each data set, as described in Section \ref{metric_calculation_details}. For each source code file in each data set, a metric file is generated in Weka ARFF format that contains STD, STN metrics, and the defectiveness class (0 or 1). The similarity metric set generated from the similarity output of MOSS is denoted as $M_{m}$ for each data set. 

\item As alternative similarity detection methods, CPD and CCFinder clone detection techniques are used to find the similarities among the files of each project. Using the similarities detected by CPD and CCFinder, the previous three steps are followed to generate alternative metric files that include STD, STN, and defectiveness (0 or 1). The metric sets generated from the similarities detected by CPD and CCFinder are denoted as $M_{c}$ and $M_{f}$ respectively. 

\item How defectiveness prediction performance changes when the sizes of the source code files are considered while calculating the suggested similarity metrics? To answer this question, the similarities detected by MOSS are used to calculate the STDS and STNS metrics that take into account file size for each file in each data set. The similarity metric set that considers the sizes of the source code files is denoted as $M_{ms}$.
\end{enumerate}

\begin{figure}[h!]
\centering
\includegraphics[width=\linewidth]{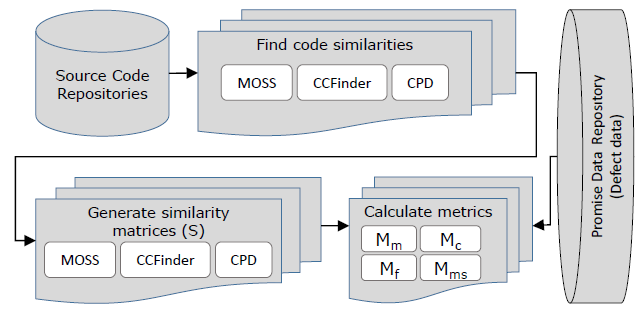}
\caption{A brief summary of the metric generation process.}
\label{fig:met_gen}
\end{figure}

The big picture of the metric extraction process is illustrated in Figure \ref{fig:met_gen} where four metric sets \ie $M_{m}$, $M_{c}$, $M_{f}$, and $M_{ms}$ are calculated for each data set based on the similarities detected by MOSS, CPD, and CCFinder.

\subsection{Experiment Design}
\subsubsection{Data Sets}

Menzies et al. state that to measure the improvement in a prediction model, a well-defined baseline should exist and that baseline needs to be based on a public domain data set and the model should be repeatable \cite{Menzies01}. The on-line Promise \cite{promiserepo} data repository is one of the largest repositories for Software Engineering research data and includes a variety of publicly available research data sets. We use open source data sets in the Promise data repository and believe that everybody else can repeat our experiments on the same data sets to verify our observations.

\begin{table}[h!]
\begin{center}
\caption{Brief summary of 10 data sets used in the experiments \label{data_sets}}{
\begin{tabular}{llrr}
\hline
Data Set & Version & \# of Instances & Defective Instances \% \\ \hline
ant & 1.7 & 730 & 22.7 \\ 
camel & 1.0 & 195 & 6.66 \\ 
ivy & 2.0 & 352 & 11.36 \\
jedit & 4.0 & 292 & 25.68 \\ 
lucene & 2.4 & 329 & 61.39 \\ 
poi & 3.0 & 437 & 64.07 \\ 
synapse & 1.2 & 255 & 33.72 \\ 
tomcat & 6.0 & 769 & 9.75 \\ 
velocity & 1.5 & 213 & 66.66 \\ 
xalan & 2.6 & 789 & 48.92 \\ 
\hline
\end{tabular}}
\end{center}
\end{table} 

To select data sets for the experiments we have two objectives. First, the data sets need to be open source since we need to measure the similarity of the files in each data set to calculate the proposed similarity metrics. Second, the data sets need to be large enough \ie include sufficient number of files to be able to perform cross validation. Based on these two objectives, we select 10 publicly available projects \cite{promise_new} \ie ant, camel, ivy, jedit, lucene, poi, synapse, tomcat, velocity, and xalan that are listed in Table \ref{data_sets} and have static code metrics available in the Promise data repository. In order to compare the performance of the proposed similarity metrics with the static code measures, the static code metrics in the Promise data repository are used. The same set of static code metrics in the repository (under the CK Metrics group) are used for all data sets. The names and sources of these static code metrics are listed in Table \ref{static_code_metrics}. The static code metric set includes object oriented metrics proposed by Chidamber and Kemerer \cite{Chidamber01}, Bansiya and Davis \cite{Bansiya01}, and the complexity metrics proposed by McCabe  \cite{McCabe02}. These object oriented and complexity metrics were explained in Section \ref{studies_sw_metrics}. The static code metric set also includes object oriented metrics defined by Schanz and Izurieta \cite{Schanz01}, Henderson-Sellers \cite{Henderson01}, and Tang et al. \cite{Tang01}.

\begin{table*}[h!]
\centering

\caption{List of static code metrics used in the Promise data repository defect data sets. \label{static_code_metrics}}{
\begin{tabular}{lll}
\hline
\textbf{Metric } & \textbf{Metric full name} & \textbf{Source} \\ \hline \hline
\textbf{wmc} & Weighted method per class & \cite{Chidamber01} \\ 
\textbf{dit} & Depth of inheritance tree & \cite{Chidamber01} \\ 
\textbf{noc} & Number of children & \cite{Chidamber01} \\ 
\textbf{cbo} & Coupling between Objects  & \cite{Chidamber01} \\ 
\textbf{rfc} & Response for class  & \cite{Chidamber01} \\ 
\textbf{lcom} & Lack of cohesion of methods  & \cite{Chidamber01} \\ 
\textbf{ca} & Afferent couplings & \cite{Schanz01} \\ 
\textbf{ce} & Efferent coupling & \cite{Schanz01} \\ 
\textbf{npm} & Number of public methods  & \cite{Bansiya01} \\ 
\textbf{lcom3} & Lack of cohesion in methods & \cite{Henderson01} \\ 
\textbf{loc} & Lines of code &  \\ 
\textbf{dam} & Data access metric & \cite{Bansiya01} \\ 
\textbf{moa} & Measure of aggregation & \cite{Bansiya01} \\ 
\textbf{mfa} & Measure of functional abstraction  & \cite{Bansiya01} \\ 
\textbf{cam} & Cohesion among methods of class & \cite{Bansiya01} \\ 
\textbf{ic} & Inheritance coupling  & \cite{Tang01} \\ 
\textbf{cbm} & Coupling between methods & \cite{Tang01} \\ 
\textbf{amc} & Average method complexity & \cite{Tang01} \\ 
\textbf{max\_cc} & max. (McCabe's cyclomatic complexity) & \cite{McCabe02} \\ 
\textbf{avg\_cc} & avg. (McCabe's cyclomatic complexity) & \cite{McCabe02} \\ \hline
\end{tabular}}

\end{table*} 

\subsubsection{Classifiers}
\label{Classifiers}
The classifiers used in this study are Random Forest (RF), Naive Bayes (NB), and $k$-Nearest Neighbours (kNN). These classifiers are selected for their common use in the software engineering defect prediction literature. In general, default settings of these classifiers are used as specified in Weka \cite{Weka01}. 

RF is an ensemble learning classification technique and one of the most successful defect prediction methods that constructs decision trees during training and outputs the class label that is the mode of the classes of the individual trees in the training set \cite{rf_Breiman}. RF is a way of averaging multiple decision trees and boosts the performance of the classification model in spite of causing a small increase in the bias.

NB is a simple probabilistic classification method that makes use of Bayes rule of conditional probability. It is named naive, because it assumes conditional independence of its attribute set. NB classifiers are successful, even on real-world data where the variables are not conditionally independent \cite{domingos_nb}. Domingos and Pazzani state that since it has advantages in terms of simplicity, learning speed, classification speed, storage space and incrementality, the use of Bayesian classifier should be considered more often \cite{domingos_nb}. 

$k$-nearest neighbor (kNN) is a classification algorithm which is also known as IBK (instance based classifier) \cite{Aha_Ibk}. In kNN an instance is labelled considering the majority label of its $k$ neighbours where $k>=1$. We use the default parameters of kNN in Weka, except we set \emph{distanceWeighting} parameter to \emph{by 1/distance} and the \emph{crossValidate} parameter to \emph{true}. Because we want to assign higher weights to the nearer (more similar) instances while deciding on the defect proneness of a specific instance in each data set, we set the \emph{distanceWeighting} parameter to \emph{1/distance}.  

\subsubsection{Performance Metric}
Considering defect prediction as a binary classification problem, let S = $\left \{ (\mathbf{x_i},y_i) \right \}$ be a set of $N$ examples where $\mathbf{x_i}$ represents software metrics and $y_i$ represents the target class \ie whether the software is defective or not. Let us assume that defective modules (positive instances) are represented with + and non defective modules (negative instances) are represented with -, so we have $y_i \in  \left \{ +,- \right \}$. Then a defect predictor function is defined as $f(x):\left \{ \mathbf{x_i}\rightarrow \left \{ +,- \right \} \right \}$.
Depending on the outcome of the defect predictor, there are four possible cases:
\begin{enumerate}
    \item \textit{True positive (TP)}: If a file is defective and is classified as defective.
    \item \textit{False negative (FN)}: If it is defective but is classified as non defective.
    \item \textit{True negative (TN)}: If it is non defective and is classified as non defective.
    \item \textit{False positive (FP)}: If it is non defective but is classified as defective
\end{enumerate}

A sample confusion matrix showing the prediction outcomes together with the actual class is shown in Table \ref{tab:ConfMatrix}.
\begin{table}[h!]
\begin{center}

\caption{A sample confusion matrix. \label{tab:ConfMatrix}}
{\begin{tabular}{l|ll}
 & \multicolumn{2}{c}{Predicted Class} \\ \hline
Actual Class & + & - \\ \hline
+ & TP & FN \\ \hline
- & FP & TN \\ \hline
\end{tabular}}
\end{center}
\end{table}
It is desirable for a defect predictor to be able to mark as many defective modules as possible while avoiding false classifications FN and FP.

Some classifiers are evaluated using their TP Rate (also called sensitivity)
\begin{equation}\ TP Rate = \frac{TP}{TP+FN} \end{equation}
or FP rate (also called false alarm rate)
\begin{equation}\ FP Rate = \frac{FP}{FP+TN} \end{equation}

It is a common way to look at the error rates of classifiers while making comparisons. However, this is not true in real life, because first, the proportions of classes are not equal. For instance, in defect prediction, the proportion of defective files are much smaller compared to the non defective ones. Furthermore, the cost of FP and FN are not the same \ie FN is more costly than FP. Considering a defective file as non defective is more costly than considering a non defective file as defective. As a result, instead of making comparisons using sensitivity or false alarm rate, it is better and more convenient to consider them together. ROC analysis is used in the literature and considers TP rate (sensitivity) and FP rate (hit) together. ROC curve is a two dimensional graphical representation where sensitivity (TP) is the y-axis and false alarm rate (FP) is the x-axis. It is desirable to have high sensitivity and small false alarm rate. So, as the area under the ROC curve (AUC) gets larger, the classifier gets better. Therefore, we use AUC as a comparison measure to compare the performance of the proposed metrics and the existing static code metrics in the literature.
 
\subsubsection{Experiment Methodology}
For each data set we compare the performance of the static code metrics and the proposed similarity metrics in terms of AUC. All classifications are carried out on the Weka experimenter with $5\times5$ folds cross validation. Static code metrics listed in Table \ref{static_code_metrics} are used from the Promise data repository. The proposed similarity metrics are calculated for each file in each data set as it is described in Section \ref{Metric Extraction Process}. The experiments carried out on each data set are:
\begin{enumerate}
\item We use the static code metrics listed in Table \ref{static_code_metrics} (denoted as $M_{sc}$) to learn the relationship between static code metrics and defect proneness for each data set.

\item The metric set based on MOSS ($M_{m}$) is used to learn the relationship of the proposed similarity metrics (STD, STN) and defect proneness for each data set.

\item To see how the results change when different similarity detection methods are used,  metric sets generated with the CPD ($M_{c}$) and CCFinder ($M_{f}$) are used to learn the relationship of the similarity metrics (STD, STN) and defect proneness for each data set.

\item Furthermore, to see if the sizes of the source code files matter while calculating the similarity metrics, the metric set (shown as $M_{ms}$ in Section \ref{Metric Extraction Process}) is used to learn the relationship of STDS and STNS metrics and defectiveness for each data set.

\item The classification techniques listed in Section \ref{Classifiers} are used to compare the results for the five metric sets defined \ie $M_{sc}$, $M_{m}$, $M_{c}$, $M_{f}$, and $M_{ms}$, too observe how the results change when different classification methods are used.

\end{enumerate}

\begin{figure}[h!]
\centering
\includegraphics[width=\linewidth]{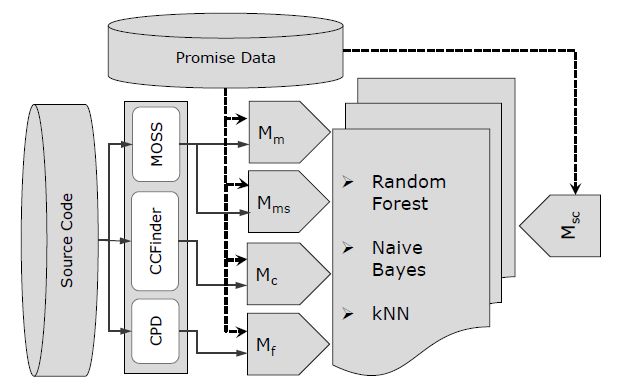}
\caption{An overview of the experiments carried out to compare the predictive power of static code and similarity metrics.}
\label{fig:exp_meth}
\end{figure}

A brief overview of the experiments is shown in Figure \ref{fig:exp_meth} where five metric sets are compared using Random Forest, Naive Bayes, and $k$-Nearest Neighbour classifiers.

\subsection{Results}
Random Forest (RF), Naive Bayes (NB), and $k$-Nearest Neighbour (kNN) classifiers explained in Section \ref{Classifiers} are used to compare the performance of the similarity metrics based on MOSS ($M_{m}$), CPD ($M_{c}$), CCFinder ($M_{f}$), and the similarity metrics based on MOSS that take file size into account ($M_{ms}$) with the static code metrics ($M_{sc}$) listed in Table \ref{static_code_metrics}. To compare the performance of different metric sets, a two tailed $t$-test with a $p$-value of 0.05 is used in all statistical significance tests for all experiments.

\subsubsection{Data Analysis}
The similarity matrices (Equation \ref{Eq:smatrix}) based on the plagiarism or clone detection techniques are sparse for all data sets. The number of rows and columns ($RC$), the number of non zero entries ($NNE$) and the density ($\rho$) of each similarity matrix generated for each data set is shown in Table \ref{tbl_sim_matrices} for MOSS, CPD, and CCFinder. The density column ($\rho$) shows the percentage of the non zero entries among the total number of entries in each similarity matrix. For instance, for camel data set, all similarity matrices have a size of 195 (195 rows and columns), and they have 2624, 46, and 21 non zero entries for MOSS, CPD, and CCFinder respectively. The densities of these similarity matrices are 6.9 \%, 0.12 \% and 0.06 \% for MOSS, CPD, and CCFinder. The similarity matrix generated with CPD is 57.04 times more sparse (2624 / 46) compared to the similarity matrix generated with MOSS. 

The similarity matrices generated from CPD are sparser compared to the similarity matrices generated from MOSS where all CPD matrices have a density value of less than 1. Except the poi and synapse data sets, the similarity matrices generated from CCFinder are also sparser compared to MOSS. For highly imbalanced data sets like camel this sparseness could be a problem if the number of non-zero similarity metrics is not enough for the classifier to learn the relationship of the similarity metrics and the defect proneness for the defective instances.

\begin{table}[h!]
\centering
\caption{For each data set, the number of rows and columns ($RC$), the number of non zero entries ($NNE$), and the percentage of non zero entries ($\rho$) in the similarity matrices are shown for MOSS, CPD, and CCFinder.}
\label{tbl_sim_matrices}
\begin{tabular}{|l|c|c|c|c|c|c|c|}
\hline
\textbf{}           & \textbf{} & \multicolumn{2}{c|}{\textbf{MOSS}} & \multicolumn{2}{c|}{\textbf{CPD}} & \multicolumn{2}{c|}{\textbf{CCFinder}} \\ \hline
\textbf{Data Sets}            & $RC$      & $NNE$           & $\rho$         & $NNE$          & $\rho$         & $NNE$             & $\rho$           \\ \hline
\textbf{ant}      & 730       & 4962          & 0.93               & 1116         & 0.21               & 921             & 0.17                 \\ 
\textbf{camel}    & 195       & 2624          & 6.90               & 46           & 0.12               & 21              & 0.06                 \\ 
\textbf{ivy}      & 352       & 5119          & 4.13               & 370          & 0.30               & 196             & 0.16                 \\ 
\textbf{jedit}    & 292       & 2592          & 3.04               & 456          & 0.53               & 191             & 0.22                 \\ 
\textbf{lucene}   & 329       & 2653          & 2.45               & 266          & 0.25               & 182             & 0.17                 \\ 
\textbf{poi}      & 437       & 7710          & 4.04               & 1538         & 0.81               & 3874            & 2.03                 \\ 
\textbf{synapse}  & 255       & 768           & 1.18               & 458          & 0.70               & 1248            & 1.92                 \\ 
\textbf{tomcat}   & 769       & 3616          & 0.61               & 1358         & 0.23               & 996             & 0.17                 \\ 
\textbf{velocity} & 213       & 2770          & 6.11               & 250          & 0.55               & 173             & 0.38                 \\ 
\textbf{xalan}    & 789       & 5259          & 0.84               & 2174         & 0.35               & 3310            & 0.53                 \\ \hline
\end{tabular}
\end{table}

The key novelty of this research is based on the hypothesis that a defective file would be more similar to defective files. In order to check that and have a better insight about the distinctive characteristics of the detected similarities, we analyze the distribution of similarities found between defective files and the distribution of similarities seen between non-defective files. Therefore, in addition to the overall number of non-zero entries in the similarity matrices, we count the number of non-zero entries where both files are defective ($E_D$), and the number of non-zero entries where both files are non-defective ($E_N$). Table \ref{similarity_of_def_and_nondef_files} shows that the majority of the detected similarities are either between defective files or non-defective files in each data set. In Table \ref{similarity_of_def_and_nondef_files}, the $S$ column shows the total percentage of similarities represented by $E_D + E_N$ for each similarity detection technique. For instance, for the ant data set, the total percentage of similarities between defective or non-defective files are 63.93 \%, 70.07 \% and 76.11 \% respectively for MOSS, CPD and CCFinder. Having a high $S$ means that defective files are more similar to the defective files and non-defective files are more similar to the non-defective files. We observe that $S$ is larger than 70 \% for 5, 7, and 10 data sets for MOSS, CPD and CCFinder respectively. In fact, for some data sets like camel, lucene, poi, and xalan, $S$ is larger than 80 \% for all similarity detection tools \ie MOSS, CPD, and CCFinder. 

\begin{table*}[h!]
\centering
\caption{The number and percentage of similarities found between defective and non-defective files.}
\label{similarity_of_def_and_nondef_files}
\begin{tabular}{|l|l|l|l|l|l|l|l|l|l|}
\hline
\textbf{}         & \multicolumn{3}{c|}{\textbf{MOSS}}               & \multicolumn{3}{c|}{\textbf{CPD}}                & \multicolumn{3}{c|}{\textbf{CCFinder}}               \\ \hline
\textbf{Data Sets} & $E_D$ & $E_N$ & $S$ & $E_D$ & $E_N$ & $S$ & $E_D$ & $E_N$ & $S$ \\ \hline
ant               & 2032            & 1140            & 63.93        & 212             & 570             & 70.07        & 182             & 519             & 76.11        \\ 
camel             & 38              & 2136            & 82.85        & 0               & 44              & 95.65        & 3               & 18              & 100.00       \\ 
ivy               & 688             & 2121            & 54.87        & 48              & 148             & 52.97        & 48              & 98              & 74.49        \\ 
jedit             & 798             & 780             & 60.88        & 66              & 214             & 61.40        & 57              & 94              & 79.06        \\ 
lucene            & 1980            & 153             & 80.40        & 182             & 40              & 83.46        & 149             & 17              & 91.21        \\ 
poi               & 6962            & 336             & 94.66        & 1282            & 120             & 91.16        & 3538            & 134             & 94.79        \\ 
synapse           & 424             & 194             & 80.47        & 176             & 104             & 61.14        & 533             & 467             & 80.13        \\ 
tomcat            & 525             & 1677            & 60.90        & 82              & 930             & 74.52        & 79              & 745             & 82.73        \\ 
velocity          & 1424            & 368             & 64.69        & 138             & 48              & 74.40        & 161             & 8               & 97.69        \\ 
xalan             & 4403            & 356             & 90.49        & 1046            & 572             & 74.43        & 2995            & 201             & 96.56        \\ \hline
\end{tabular}
\end{table*}

\subsubsection{Analysis of Results with RF}
Using the RF classifier, the AUC values obtained for each data set is shown in Table \ref{tbl_AUC_RF}. We first observe that $M_{m}$ metrics are better defect predictors compared to the static code metrics ($M_{sc}$) for all data sets. The AUC values of $M_{m}$ are consistently and clearly higher than the AUC values of $M_{sc}$ in all data sets and the differences are statistically significant except ivy and velocity data sets. Furthermore, $M_{m}$ metrics are  better than other similarity metric sets \ie  $M_{c}$ and  $M_{f}$ for all data sets and the differences are statistically significant.
\begin{table}[h!]
\begin{center}
\caption{The AUC values of the studied metric sets with RF classifier. \label{tbl_AUC_RF}}{
\begin{tabular}{llllll}
\hline
\textbf{Data Sets} & $M_{sc}$ & $M_{m}$& $M_{c}$& $M_{f}$& $M_{ms}$ \\  \hline
\textbf{ant} & 0.73 & \textbf{0.96} & 0.73 & 0.74 & 0.90 \\
\textbf{camel} & 0.65 & \textbf{0.81} & 0.25 & 0.62 & 0.77 \\
\textbf{ivy} & 0.75 & \textbf{0.83} & 0.63 & 0.80 & 0.85 \\
\textbf{jedit} & 0.73 & \textbf{0.90} & 0.60 & 0.75 & 0.86  \\
\textbf{lucene} & 0.61 & \textbf{0.95} & 0.84 & 0.70 & 0.90 \\
\textbf{poi} & 0.83 & \textbf{0.89} & 0.77 & 0.78 & 0.86  \\ 
\textbf{synapse} & 0.69 & \textbf{0.96} & 0.76 & 0.76 & 0.90 \\ 
\textbf{tomcat} & 0.72 & \textbf{0.93} & 0.51 & 0.72 & 0.89 \\
\textbf{velocity} & 0.79 & \textbf{0.87} & 0.78 & 0.70 & 0.87 \\
\textbf{xalan} & 0.75 & \textbf{0.94} & 0.73 & 0.79 & 0.89 \\ \hline
\end{tabular}}
\end{center}

\end{table}

Furthermore, we observe that similarity metrics calculated using the CPD ($M_{c}$) clone detection tool are not better defect predictors compared to the static code metrics ($M_{sc}$). Because, although the AUC values of $M_{c}$ are higher compared to the AUC values of static code metrics for lucene and synapse data sets, $M_{c}$ metric set has lower or comparable AUC values for other data sets. Similarly, except the ivy, lucene, synapse, and xalan data sets, the AUC value of the $M_{f}$ metric set is comparable or lower when compared to the static code metrics. The similarity matrices generated with CPD and CCFinder are more sparse compared to the similarity matrices generated with MOSS. The amount of similarity detected by CPD and CCFinder is less compared to MOSS, because CPD and CCFinder focus on the syntactic similarity and detect source code clones only. Therefore, for the majority of files in each data set the similarity metrics generated from CPD and CCFinder are zero, and that might be degrading the performance of the prediction model (See the lower AUC value of $M_{c}$ and  $M_{f}$ for the highly imbalanced camel data set which also has a very sparse similarity matrix). 

Moreover, we observe that the similarity metrics that take the file size into account ($M_{ms}$) are better defect predictors when compared to the static code metrics and the differences are significant in all data sets except poi and velocity. However, when the performances of $M_{m}$ and $M_{ms}$ are compared, we observe that $M_{ms}$ achieves a slightly higher AUC in ivy data set. But, $M_{m}$ metrics are better defect predictors for the remaining data sets and the differences in the AUC values are statistically significant except poi and velocity data sets. Therefore, we conclude that taking the file size into account does not make a significant contribution to the prediction model.

\begin{figure}[h!]
\centering
\includegraphics[width=\linewidth]{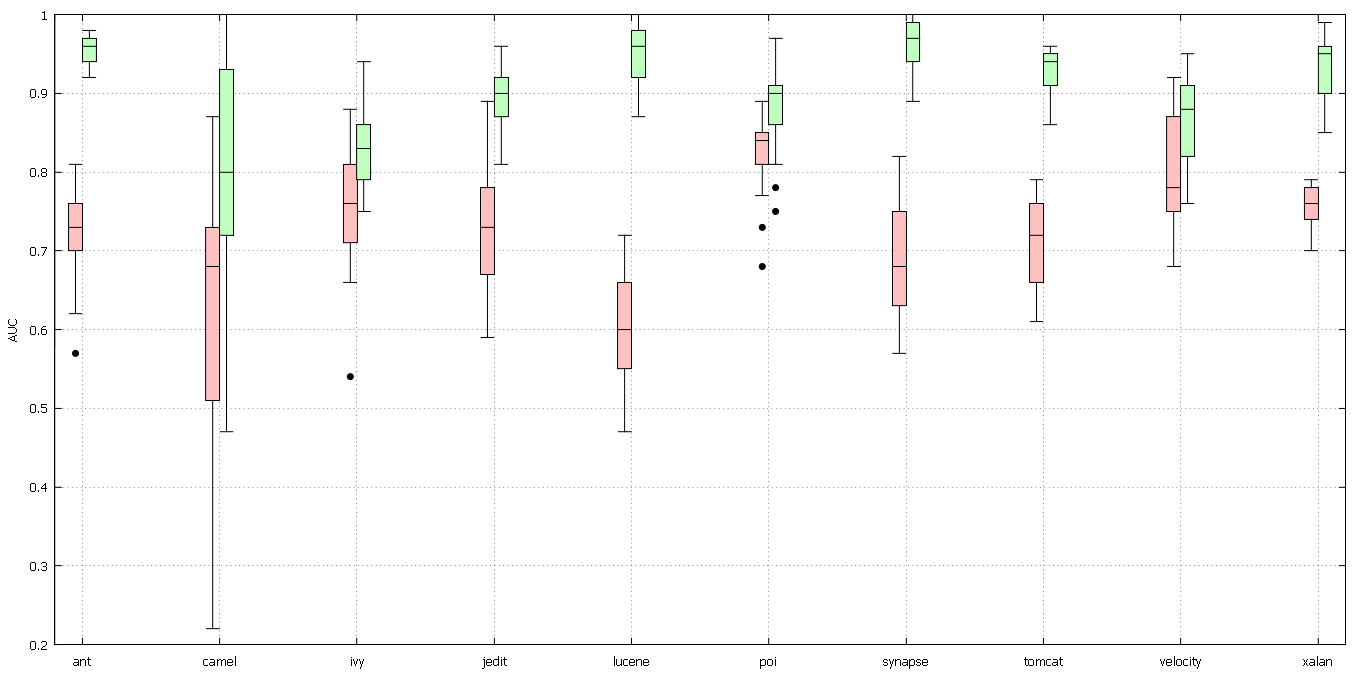}
\caption{Box plots of AUC values with RF classifier for \scm\  (\crule[flamingopink]{0.2cm}{0.2cm}) and \sm\  (\crule[etonblue]{0.2cm}{0.2cm}).}
\label{fig:boxplotrf}
\end{figure}

As a result, when the RF classifier is used $M_m$ metric set achieves a better performance when compared to other similarity metric sets $M_{c}$, $M_{f}$, and $M_{ms}$. A comparison of the AUC values of $M_{m}$ and $M_{sc}$ is shown in the box plot in Figure \ref{fig:boxplotrf}. The AUC values of $M_{m}$ are consistently better in all data sets when compared to the AUC values of $M_{sc}$ except ivy and velocity. 
 
\subsubsection{Analysis of Results with NB}
\begin{figure}[h!]
\centering
\includegraphics[width=\linewidth]{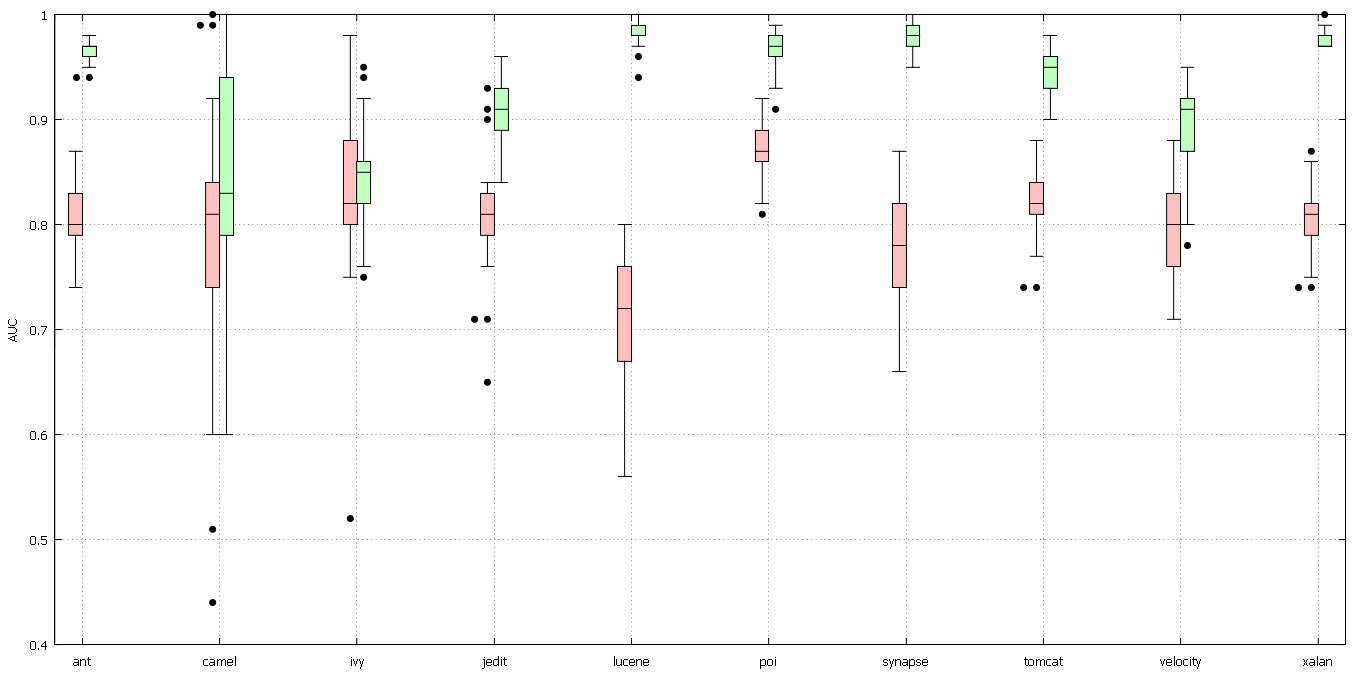}
\caption{Box plots of AUC values with NB classifier for \scm\  (\crule[flamingopink]{0.2cm}{0.2cm}) and \sm\  (\crule[etonblue]{0.2cm}{0.2cm}).}
\label{fig:boxplotnb}
\end{figure}

Naive Bayes classifier (NB) is used to cross check the results found with Random Forest (RF). The AUC values of all metric sets \ie $M_{sc}$, $M_{m}$, $M_{c}$, $M_{f}$, and $M_{ms}$ are shown in Table \ref{table_AUC_NB}. The AUC values observed for the similarity metrics calculated with MOSS ($M_{m}$) are higher when compared to the AUC values found for the static code metrics ($M_{sc}$). Furthermore, in eight of the ten data sets (except camel and ivy) the differences in the AUC values are statistically significant. The box plots of the AUC values for $M_{m}$ and $M_{sc}$ are shown in Figure \ref{fig:boxplotnb} for all data sets. With the NB classifier, the following observations support previous findings with RF.
\begin{table}[h!]
\begin{center}
\caption{The AUC values of different metrics with NB. \label{table_AUC_NB}}{
\begin{tabular}{llllll}
\hline
\textbf{Data Sets} & $M_{sc}$ & $M_{m}$& $M_{c}$& $M_{f}$& $M_{ms}$ \\  \hline
\textbf{ant} & 0.80 & \textbf{0.97}  & 0.77 & 0.74 & 0.94 \\ 
\textbf{camel} & 0.81 & \textbf{0.84} & 0.25 & 0.59 & 0.82 \\ 
\textbf{ivy} & 0.82 & \textbf{0.85} & 0.71 & 0.78 & 0.84 \\ 
\textbf{jedit} & 0.81  & \textbf{0.91} & 0.7  & 0.73 & 0.88 \\ 
\textbf{lucene} & 0.72 & \textbf{0.98} & 0.85 & 0.72 & 0.94 \\ 
\textbf{poi} & 0.87 & \textbf{0.97}  & 0.80 & 0.73 & 0.97 \\ 
\textbf{synapse} & 0.78 & \textbf{0.98} & 0.74 & 0.78 & 0.94  \\ 
\textbf{tomcat} & 0.82 & \textbf{0.94}  & 0.71 & 0.72 & 0.94 \\ 
\textbf{velocity} & 0.80 & \textbf{0.90} & 0.85 & 0.71 & 0.87 \\ 
\textbf{xalan} & 0.81 & \textbf{0.98} & 0.80 & 0.75 & 0.98 \\ \hline
\end{tabular}}
\end{center}

\end{table}
\begin{itemize}
\item The similarity matrices generated from the CPD and CCFinder are sparse compared to the similarity matrices generated from MOSS, therefore $M_{c}$ and $M_{f}$ metric sets are not successful defect predictors compared to $M_{m}$. Moreover, $M_{c}$ and $M_{f}$ metric sets have comparatively poor performance for the highly imbalanced camel data set that has a very sparse similarity matrix for CPD and CCFinder. (See the densities of the similarity matrices for MOSS, CPD, and CCFinder in Table \ref{tbl_sim_matrices}).
\item Taking file size into account does not improve the defect prediction model, since the AUC values of $M_{ms}$ are lower compared to the AUC values of $M_{m}$ in 7 of 10 data sets (See Table \ref{table_AUC_NB}).
\end{itemize}

\subsubsection{Analysis of Results with kNN}
\begin{figure}[h!]
\centering
\includegraphics[width=\linewidth]{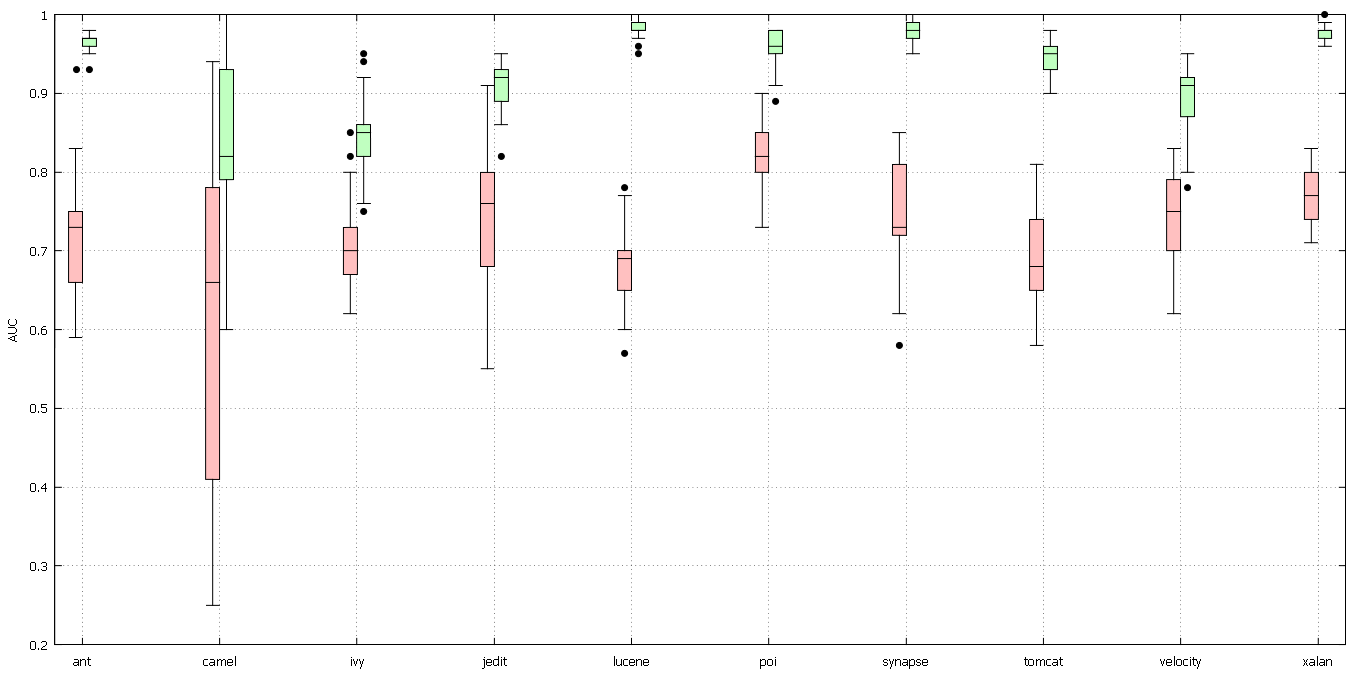}
\caption{Box plots of AUC values with KNN classifier for \scm\  (\crule[flamingopink]{0.2cm}{0.2cm}) and \sm\  (\crule[etonblue]{0.2cm}{0.2cm}).}
\label{fig:boxplotknn}
\end{figure}

The $k$-Nearest Neighbour algorithm (kNN) is used to compare the performance of the calculated similarity metrics ($M_{m}$, $M_{c}$, $M_{f}$, and $M_{ms}$) with the static code metrics ($M_{sc}$). The AUC values found with the kNN classifier is shown in Table \ref{table_AUC_KNN} for each data set. We observe that the AUC values of $M_{m}$ are higher than the AUC values of $M_{sc}$ in all data sets and all the differences are statistically significant. The statistically significant differences between the AUC values of $M_{m}$ and the static code metrics ($M_{sc}$) are shown in the box plot in Figure \ref{fig:boxplotknn}.

\begin{table}[h!]
\begin{center}
\caption{The AUC values of different metrics with kNN. \label{table_AUC_KNN}}{
\begin{tabular}{llllll}
\hline
\textbf{Data Sets} & $M_{sc}$ & $M_{m}$& $M_{c}$& $M_{f}$ & $M_{ms}$ \\  \hline
\textbf{ant} & 0.72 & \textbf{0.96} & 0.73 & 0.70 & 0.94 \\ 
\textbf{camel} & 0.63 & \textbf{0.83} & 0.46 & 0.60 & 0.82 \\ 
\textbf{ivy} & 0.71 & \textbf{0.85} & 0.61 & 0.70 & 0.80 \\ 
\textbf{jedit} & 0.75 & \textbf{0.91} & 0.59 & 0.70 & 0.88 \\ 
\textbf{lucene} & 0.68 & \textbf{0.98} & 0.88 & 0.71 & 0.94 \\ 
\textbf{poi} & 0.82 & \textbf{0.96} & 0.74 & 0.77 & 0.96 \\ 
\textbf{synapse} & 0.74 & \textbf{0.98} & 0.72 & 0.74 & 0.94  \\ 
\textbf{tomcat} & 0.69 & \textbf{0.94}  & 0.60 & 0.69 & 0.94 \\ 
\textbf{velocity} & 0.74 & \textbf{0.90} & 0.83 & 0.71 & 0.87 \\ 
\textbf{xalan} & 0.77 & \textbf{0.98} & 0.72 & 0.76 & 0.98 \\ \hline
\end{tabular}}
\end{center}
\end{table}

Similar to the previous observations with other classifiers, since the underlying similarity matrices are sparser, the similarity metrics calculated via CPD ($M_{c}$) and CCFinder ($M_{f}$) do not perform well when compared to $M_{m}$ metric set. Considering all similarity metric sets \ie $M_{m}$, $M_{c}$, $M_{f}$, and $M_{ms}$, the metric set $M_{m}$ achieves the highest performance in terms of AUC when kNN is used. Furthermore, it is hard to conclude that $M_{c}$ and $M_{f}$ are better defect predictors when compared to the static code metrics, because they both perform worse on some data sets like camel, poi, and xalan when compared to the $M_{sc}$ metric set.

Moreover, we find out that considering the size of the files while calculating the similarity metrics does not add a value to the prediction model. The AUC values of $M_{ms}$ are comparable with the AUC values of $M_{m}$ in some data sets like poi, tomcat, and xalan. However, the AUC values of $M_{m}$ are better in the remaining data sets although the differences are not significant in ant, camel, jedit, and velocity data sets. As a result, looking at the overall performance of $M_{m}$ and $M_{ms}$ in all data sets, it is hard to conclude that considering the file size (while calculating the proposed similarity metrics) improves the prediction model.  

\subsubsection{Summary of Results}
First we observe that the performance of the proposed similarity metrics is consistent for all classifiers. We show that using the similarity metrics based on the similarities detected among the files of a software project, it is possible to predict software defectiveness. The following are some common observations with different classifiers:

\begin{itemize}
\item Similarity metrics calculated from the output of the plagiarism tool ($M_{m}$) achieve a higher performance compared to the static code metrics ($M_{sc}$) in terms of the AUC. The differences in the AUC values of $M_{m}$ and $M_{sc}$ are statistically significant for all classifiers in almost all data sets.

\item $M_{m}$ metric set is better in predicting software defectiveness when compared to other similarity metrics \ie $M_{c}$ and $M_{f}$. When the underlying similarity detection method is not successful in catching existing similarities among the files in a project, then the performance of a classifier might get worse. Therefore, we observe that the similarity metric sets calculated via CPD and CCFinder do not perform well when compared to the static code metrics.

\item Considering file sizes while calculating the similarity metrics does not improve the defectiveness prediction performance of the classifiers.

\item If a data set is skewed and the density of its similarity matrix is low, a classifier might have a relatively poorer performance in predicting defective files. We observe lower AUC values for the highly skewed camel data set when the similarity matrices generated via CPD and CCFinder are used (See Tables \ref{tbl_AUC_RF}, \ref{table_AUC_NB}, and \ref{table_AUC_KNN}).  
\end{itemize}

\section{Threats to Validity}
Internal, construct and external validity are three types of validity threats that should be taken into consideration in research studies \cite{Perry:2000}. We briefly explain the measures taken to mitigate these validity threats.

An internal validity threat implies that cause effect relationships between the dependent and independent variables are not established properly. In our research context, an internal validity threat might be related to the selection of classification methods and their parameters, data sets, and the underlying similarity detection tools. To mitigate the internal validity threats, first we use different common classifiers with no parameter tuning and apply $5\times5$ folds cross validation during each experiment. Moreover, we use 10 different publicly available data sets to show that proposed similarity metrics have much better defectiveness prediction capability compared to the traditional static code metrics. But, we should note that the similarity metrics proposed in this study are highly dependent on the underlying similarity detection technique. If the amount of detected similarity is low (generated similarity matrix is so sparse), then the similarity metrics calculated from such a matrix could have a poorer defect prediction performance compared to the existing software metrics. 

Construct validity threats can be observed when there are errors in the measurements. To mitigate construct validity, we automate the similarity metric calculation to prevent any calculation errors. Furthermore, we use different measurement tools to calculate source code similarities among the files of the data sets.

An external validity threat implies that the results observed on one data set are not valid for others. Although we use 10 open source data sets, alternative classifiers, and $5\times5$ folds cross validation in our experiments, it is not easy to generalize our findings. Our study is replicable since anyone else can calculate the same metrics using the same data sets to cross check our observations. We believe that further studies with different data sets and different classifiers are needed to justify our findings. 

\section{Conclusion}
\label{Conclusion}
Source code similarity has been applied in software engineering literature to detect code clones or check for software plagiarism. However, there is no prior study using source code similarity metrics for defect prediction purposes. This paper proposes to use two novel software metrics which are based on the source code similarity to predict defect proneness. While one of the proposed metrics measures the overall similarity of a file to the remaining files within a software project that are marked as defective (STD), the other metric measures its overall similarity to the files that are marked as non defective (STN). Commonly used defect prediction methods \ie Random Forest, Naive Bayes and $k$-Nearest Neighbour classifiers are used with 5 $\times$ 5 folds cross validation, to check the predictive power of the proposed similarity metrics on 10 different open source data sets. We show that depending on the amount of detected similarity, proposed source code similarity metrics are better defect predictors compared to the traditional static code metrics in terms of the area under the ROC curve (AUC). Our results strongly endorse the usage of source code similarity metrics for software defect prediction.   

Furthermore, we also check how the prediction capability of the proposed similarity metrics changes when the size of the files are taken into account while calculating these metrics. We show that including the file size in the metric calculation process does not add a significant value to the prediction model in none of the used projects.

Moreover, proposed source code similarity metrics are based on the amount of the similarity detected among the files of a software system. Therefore, we use both plagiarism and clone detection techniques to detect similarities within each software project and generate alternative source code similarity matrices for each data set. We observe that the performance of the proposed similarity metrics can increase or decrease depending on the amount of similarity detected by the underlying similarity detection technique. We show that the similarity matrices calculated using the MOSS plagiarism tool are more dense compared to the similarity matrices generated from the output of the clone detection techniques \ie CPD and CCFinder. For all data sets and with all classifiers, we observe that source code similarity metrics that are based on a dense similarity matrix (MOSS) are better defect predictors compared to the similarity metrics derived from a less dense (more sparse) similarity matrix.

There might be several reasons to have similar pieces of code or clones in software projects. First, copy and pasting a piece of code in the same project is a bad practice. Because, the code that is copied should be included in a method or class and called repeatedly. However, most of the time copy and pasting cannot be prevented even in the enterprise level software projects. Second, each developer has a certain source code development style like following certain custom coding patterns, and most of the time this style is consistent within a project or even across projects. One may expect to have a higher amount of similarity among the source code files developed by the same developer. Therefore, it is possible to find clones or very similar code pieces in almost all software projects. On the other hand, if the amount of similarity among the files of a software system is high, one might expect defective files to be more similar to other defective files. Similarly, non-defective files are expected to be more similar to other non-defective files. Because, having similar code pieces or common clones among different source code files, increases the probability of sharing a defective or bug free piece of code. In fact, this research is showing that if there is a high amount of similarity in a software project, the overall similarity of each file to the defective and non-defective files in the project could be used to predict defectiveness. 
 
The Halstead and McCabe static code metrics were defined around 1970s and are still used in the defect prediction literature. Although there have been numerous phenomenal advances in the compiler, software development, and distributed computing technologies since 1970, relatively there are not enough studies regarding inter module software defect prediction metrics. Proposed metrics measure the inter module characteristics within a software project and have shown to be better defect predictors compared to the traditional static code metrics. As software projects get larger and the amount of common source code among software modules increases, more studies that focus on the inter module features of software projects are needed to achieve better defect prediction results. 

This paper uses open source projects to extract similarity metrics and compares the performance of these metrics with the static code metrics which are available in the Promise data repository. Therefore, the study is reproducible \ie anybody else might use the same set of projects and the static code metrics in the Promise data repository, iterate through the defined experiment steps to generate similar results. Moreover, the study presented in this paper is also replicable. Because, the methodology defined in this study could easily be expanded to use a new data set other than the Promise data repository that might include a different set of software metrics and defect data.   

As a future work, we plan to revise our study to merge the similarity matrices calculated using different similarity, plagiarism or clone detection techniques, recalculate the similarity metrics based on a composite matrix and check if the prediction capability of the similarity measures changes. As the amount of the similarity detected among the files of a software system gets higher, the proposed similarity metrics could achieve a higher defect prediction performance.

\clearpage
%

\bibliographystyle{ieeetran}
\bibliography{references}

\begin{thebibliography}{10}

\bibitem{promiserepo}
The promise repository of empirical software engineering data.
\newblock \url{http://openscience.us/repo/defect/ck/}, 2015.

\bibitem{Aha_Ibk}
D.~W. Aha, D.~Kibler, and M.~K. Albert.
\newblock Instance-based learning algorithms.
\newblock {\em Machine Learning}, 6(1):37--66, 1991.

\bibitem{Moss01}
A.~Aiken.
\newblock Moss (measure of software similarity).
\newblock http://cs.stanford.edu/~aiken/moss/, 1997.

\bibitem{Bansiya01}
J.~Bansiya and C.~Davis.
\newblock A hierarchical model for object-oriented design quality assessment.
\newblock {\em IEEE Transactions on Software Engineering}, 28:4--17, 2002.

\bibitem{Basili:1996}
V.~R. Basili, L.~C. Briand, and W.~L. Melo.
\newblock A validation of object-oriented design metrics as quality indicators.
\newblock {\em IEEE Transactions on Software Engineering}, 22(10):751--761,
  Oct. 1996.

\bibitem{Baxter:1998}
I.~D. Baxter, A.~Yahin, L.~Moura, M.~Sant'Anna, and L.~Bier.
\newblock Clone detection using abstract syntax trees.
\newblock In {\em Proceedings of the International Conference on Software
  Maintenance}, ICSM '98, pages 368--, Washington, DC, USA, 1998. IEEE Computer
  Society.

\bibitem{Boehm:1988}
B.~W. Boehm and P.~N. Papaccio.
\newblock Understanding and controlling software costs.
\newblock {\em IEEE Transactions on Software Engineering}, 14(10):1462--1477,
  Oct. 1988.

\bibitem{rf_Breiman}
L.~Breiman.
\newblock Random forests.
\newblock {\em Machine Learning}, 45(1):5--32, 2001.

\bibitem{Brooks:1995}
F.~P. Brooks, Jr.
\newblock {\em The Mythical Man-month (Anniversary Ed.)}.
\newblock Addison-Wesley Longman Publishing Co., Inc., Boston, MA, USA, 1995.

\bibitem{Chidamber01}
S.~R. Chidamber and C.~F. Kemerer.
\newblock Towards a metrics suite for object oriented design.
\newblock {\em SIGPLAN}, 26(11):197--211, 1991.

\bibitem{Cosma01}
G.~Cosma.
\newblock {\em An approach to source-code plagiarism detection investigation
  using latent semantic analysis}.
\newblock PhD thesis, University of Warwick, Coventry CV4 7AL, UK, 2008.

\bibitem{DAmbros2012}
M.~D'Ambros, M.~Lanza, and R.~Robbes.
\newblock Evaluating defect prediction approaches: a benchmark and an extensive
  comparison.
\newblock {\em Empirical Software Engineering}, 17(4):531--577, Aug 2012.

\bibitem{Daskalantonakis}
M.~K. Daskalantonakis.
\newblock A practical view of software measurement and implementation
  experiences within motorola.
\newblock {\em IEEE Transactions on Software Engineering}, 18(11):998--1010,
  1992.

\bibitem{domingos_nb}
P.~Domingos and M.~Pazzani.
\newblock On the optimality of the simple bayesian classifier under zero-one
  loss.
\newblock {\em Machine Learning}, 29(2):103--130, 1997.

\bibitem{Fagan1976}
M.~E. Fagan.
\newblock Design and code inspections to reduce errors in program development.
\newblock {\em IBM Systems Journal}, 15(3):182 -- 211, 1976.

\bibitem{Norman03}
N.~E. Fenton and M.~Neil.
\newblock A critique of software defect prediction models.
\newblock {\em IEEE Transactions on Software Engineering}, 25:675--689, 1999.

\bibitem{Fenton:1998}
N.~E. Fenton and S.~L. Pfleeger.
\newblock {\em Software Metrics: A Rigorous and Practical Approach}.
\newblock PWS Publishing Co., Boston, MA, USA, 2nd edition, 1998.

\bibitem{Graves01}
T.~L. Graves, A.~F. Karr, J.~Marron, and H.~Siy.
\newblock Predicting fault incidence using software change history.
\newblock {\em IEEE Transactions on Software Engineering}, 26(7):653--661,
  April 2000.

\bibitem{Gyimothy01}
T.~Gyimothy, R.~Ferenc, and I.~Siket.
\newblock Empirical validation of object-oriented metrics on open source
  software for fault prediction.
\newblock {\em IEEE Transactions on Software Engineering}, 31(10):897--910,
  2005.

\bibitem{Weka01}
M.~Hall, E.~Frank, G.~Holmes, B.~Pfahringer, P.~Reutemann, and I.~H. Witten.
\newblock {The WEKA data mining software: an update}.
\newblock {\em SIGKDD Explorations}, 11(1):10--18, 2009.

\bibitem{Hall2012}
T.~Hall, S.~Beecham, D.~Bowes, D.~Gray, and S.~Counsell.
\newblock A systematic literature review on fault prediction performance in
  software engineering.
\newblock {\em IEEE Transactions on Software Engineering}, 38(6):1276--1304,
  Nov. 2012.

\bibitem{Halstead01}
M.~H. Halstead.
\newblock {\em Elements of Software Science (Operating and programming systems
  series)}.
\newblock Elsevier Science Inc., New York, NY, USA, 1977.

\bibitem{Henderson01}
B.~Henderson-Sellers.
\newblock {\em Object-oriented Metrics: Measures of Complexity}.
\newblock Prentice-Hall, Inc., Upper Saddle River, NJ, USA, 1996.

\bibitem{promise_new}
M.~Jureczko and L.~Madeyski.
\newblock Towards identifying software project clusters with regard to defect
  prediction.
\newblock In {\em Proceedings of the 6th International Conference on Predictive
  Models in Software Engineering}, PROMISE '10, pages 9:1--9:10, New York, NY,
  USA, 2010. ACM.

\bibitem{Kamiya1019480}
T.~Kamiya, S.~Kusumoto, and K.~Inoue.
\newblock Ccfinder: a multilinguistic token-based code clone detection system
  for large scale source code.
\newblock {\em IEEE Transactions on Software Engineering}, 28(7):654--670, Jul
  2002.

\bibitem{karp87}
R.~M. Karp and M.~O. Rabin.
\newblock Efficient randomized pattern-matching algorithms.
\newblock {\em IBM Journal of Research and Development}, 31:249--260, 1987.

\bibitem{Kontogiannis:1996}
K.~A. Kontogiannis, R.~Demori, E.~Merlo, M.~Galler, and M.~Bernstein.
\newblock Reverse engineering.
\newblock chapter Pattern Matching for Clone and Concept Detection, pages
  77--108. Kluwer Academic Publishers, Norwell, MA, USA, 1996.

\bibitem{Krishnan2011}
S.~Krishnan, C.~Strasburg, R.~R. Lutz, and K.~Go\v{s}eva-Popstojanova.
\newblock Are change metrics good predictors for an evolving software product
  line?
\newblock In {\em Proceedings of the 7th International Conference on Predictive
  Models in Software Engineering}, Promise '11, pages 7:1--7:10, New York, NY,
  USA, 2011. ACM.

\bibitem{Lessmann2008}
S.~Lessmann, B.~Baesens, C.~Mues, and S.~Pietsch.
\newblock Benchmarking classification models for software defect prediction: A
  proposed framework and novel findings.
\newblock {\em IEEE Transactions on Software Engineering}, 34(4):485--496, July
  2008.

\bibitem{Li1610609}
Z.~Li, S.~Lu, S.~Myagmar, and Y.~Zhou.
\newblock Cp-miner: finding copy-paste and related bugs in large-scale software
  code.
\newblock {\em IEEE Transactions on Software Engineering}, 32(3):176--192,
  March 2006.

\bibitem{Liu:2006}
C.~Liu, C.~Chen, J.~Han, and P.~S. Yu.
\newblock Gplag: Detection of software plagiarism by program dependence graph
  analysis.
\newblock In {\em Proceedings of the 12th ACM SIGKDD International Conference
  on Knowledge Discovery and Data Mining}, KDD '06, pages 872--881, New York,
  NY, USA, 2006. ACM.

\bibitem{McCabe02}
T.~McCabe.
\newblock A complexity measure.
\newblock {\em IEEE Transactions on Software Engineering}, 2:308--320, 1976.

\bibitem{Meneely:2008}
A.~Meneely, L.~Williams, W.~Snipes, and J.~Osborne.
\newblock Predicting failures with developer networks and social network
  analysis.
\newblock In {\em Proceedings of the 16th ACM SIGSOFT International Symposium
  on Foundations of Software Engineering}, SIGSOFT '08/FSE-16, pages 13--23,
  New York, NY, USA, 2008. ACM.

\bibitem{Menzies01}
T.~Menzies, J.~Greenwald, and A.~Frank.
\newblock Data mining static code attributes to learn defect predictors.
\newblock {\em IEEE Transactions on Software Engineering}, 33:2--13, 2007.

\bibitem{Moser:2008}
R.~Moser, W.~Pedrycz, and G.~Succi.
\newblock A comparative analysis of the efficiency of change metrics and static
  code attributes for defect prediction.
\newblock In {\em Proceedings of the 30th International Conference on Software
  Engineering}, ICSE '08, pages 181--190, New York, NY, USA, 2008. ACM.

\bibitem{NagappanB07}
N.~Nagappan and T.~Ball.
\newblock Using software dependencies and churn metrics to predict field
  failures: An empirical case study.
\newblock In {\em ESEM}, pages 364--373. IEEE Computer Society, 2007.

\bibitem{okutan2}
A.~Okutan and O.~Yildiz.
\newblock Software defect prediction using bayesian networks.
\newblock {\em Empirical Software Engineering}, pages 1--28, 2012.

\bibitem{Ostrand10}
T.~J. Ostrand, E.~J. Weyuker, and R.~M. Bell.
\newblock Programmer-based fault prediction.
\newblock In T.~Menzies and G.~Koru, editors, {\em PROMISE}. ACM, 2010.

\bibitem{Perry:2000}
D.~E. Perry, A.~A. Porter, and L.~G. Votta.
\newblock Empirical studies of software engineering: a roadmap.
\newblock In {\em Proceedings of the Conference on The Future of Software
  Engineering}, ICSE '00, pages 345--355, New York, NY, USA, 2000. ACM.

\bibitem{Prechelt01}
L.~Prechelt, G.~Malpohl, and M.~Philippsen.
\newblock Jplag: Finding plagiarisms among a set of programs.
\newblock Technical report, University of Karlsruhe, Department of Informatics,
  2000.

\bibitem{PurushothamanP05}
R.~Purushothaman and D.~E. Perry.
\newblock Toward understanding the rhetoric of small source code changes.
\newblock {\em IEEE Transactions on Software Engineering}, (6):511--526, 2005.

\bibitem{Ragkhitwetsagul}
C.~Ragkhitwetsagul, J.~Krinke, and D.~Clark.
\newblock Similarity of source code in the presence of pervasive modifications.
\newblock In {\em 2016 IEEE 16th International Working Conference on Source
  Code Analysis and Manipulation (SCAM)}, pages 117--126, Oct 2016.

\bibitem{Schanz01}
T.~Schanz and C.~Izurieta.
\newblock Object oriented design pattern decay: a taxonomy.
\newblock In {\em Proceedings of the 2010 ACM-IEEE International Symposium on
  Empirical Software Engineering and Measurement}, ESEM '10, New York, NY, USA,
  2010. ACM.

\bibitem{Schleimer2003}
S.~Schleimer, D.~S. Wilkerson, and A.~Aiken.
\newblock Winnowing: Local algorithms for document fingerprinting.
\newblock In {\em Proceedings of the 2003 ACM SIGMOD International Conference
  on Management of Data}, SIGMOD '03, pages 76--85, New York, NY, USA, 2003.
  ACM.

\bibitem{Schroter06}
A.~Schr�ter, T.~Zimmermann, R.~Premraj, and A.~Zeller.
\newblock If your bug database could talk.
\newblock In {\em In Proceedings of the 5th International Symposium on
  Empirical Software Engineering, Volume II: Short Papers and Posters}, pages
  18--20, 2006.

\bibitem{Shepperd01}
M.~Shepperd and G.~Kadoda.
\newblock Comparing software prediction techniques using simulation.
\newblock {\em IEEE Transactions on Software Engineering}, 27, November 2001.

\bibitem{Shin11}
Y.~Shin, A.~Meneely, L.~Williams, and J.~A. Osborne.
\newblock Evaluating complexity, code churn, and developer activity metrics as
  indicators of software vulnerabilities.
\newblock {\em IEEE Transactions on Software Engineering}, 37(6):772--787,
  2011.

\bibitem{Song2011}
Q.~Song, Z.~Jia, M.~Shepperd, S.~Ying, and J.~Liu.
\newblock A general software defect-proneness prediction framework.
\newblock {\em IEEE Transactions on Software Engineering}, 37(3):356--370, May
  2011.

\bibitem{Tang01}
M.-H. Tang, M.-H. Kao, and M.-H. Chen.
\newblock An empirical study on object-oriented metrics.
\newblock In {\em Proc. International Software Metrics Symposium}, pages
  242--249. IEEE, 1999.

\bibitem{Wahler:2004}
V.~Wahler, D.~Seipel, J.~W.~v. Gudenberg, and G.~Fischer.
\newblock Clone detection in source code by frequent itemset techniques.
\newblock In {\em Proceedings of the Source Code Analysis and Manipulation,
  Fourth IEEE International Workshop}, SCAM '04, pages 128--135, Washington,
  DC, USA, 2004. IEEE Computer Society.

\bibitem{Weyuker01}
E.~J. Weyuker, T.~J. Ostrand, and R.~M. Bell.
\newblock Do too many cooks spoil the broth? using the number of developers to
  enhance defect prediction models.
\newblock {\em Empirical Software Engineering}, 13(5):539--559, 2008.

\bibitem{Yourdon}
E.~Yourdon and L.~L. Constantine.
\newblock {\em Structured Design: Fundamentals of a Discipline of Computer
  Program and Systems Design}.
\newblock Prentice-Hall, Inc., Upper Saddle River, NJ, USA, 1st edition, 1979.

\bibitem{Zimmermann:2008}
T.~Zimmermann and N.~Nagappan.
\newblock Predicting defects using network analysis on dependency graphs.
\newblock In {\em Proceedings of the 30th International Conference on Software
  Engineering}, ICSE '08, pages 531--540, New York, NY, USA, 2008. ACM.

\end{thebibliography}
\vspace{-10 cm}
\begin{IEEEbiography}
    [{\includegraphics[width=1in,height=1.25in,clip,keepaspectratio]{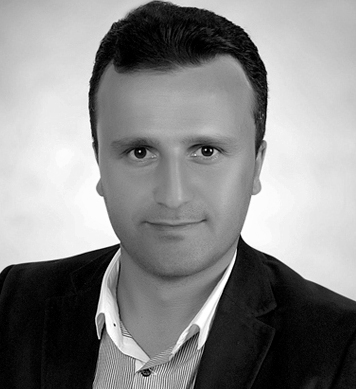}}]{Ahmet Okutan}
Dr. Ahmet Okutan received his B.S.degree in Computer Engineering from Bosphorus University, Istanbul, Turkey in 1998. He received his M.S. and Ph.D. degrees in Computer Engineering from Isik University, Istanbul, Turkey in 2002 and 2008 respectively. Dr. Okutan is currently a Post Doctoral Research Fellow in the Computer Engineering Department at Rochester Institute of Technology. He worked as analyst, architect, developer, and project manager in more than 20 large scale software projects. He has professional experience regarding software design and development, mobile application development, database management systems, and web technologies for more than 18 years. His current research interests include cyber attack forecasting, software quality prediction and software defectiveness prediction.
\end{IEEEbiography}

\end{document}